\documentclass[preprint]{aastex}

\shorttitle{Variable Stars in LMC Disk-1}
\shortauthors{Kaluzny, Mochnacki \& Rucinski}

\begin{document}

\title{Variable Stars in the Large Magellanic
Cloud: Discovery of Extragalactic W~UMa 
Binaries\footnote{Based on the data obtained at Las Campanas
Observatory, operated by the Carnegie Institution of Washington,
during the University of Toronto time allocation.}}

\author{
Janusz Kaluzny}
\affil{Copernicus Astronomical Center, Bartycka 18,
00-716 Warsaw, Poland}
\email{jka@camk.edu.pl}
\and
\author{Stefan Mochnacki, 
Slavek M. Rucinski}
\affil{David Dunlap Observatory\\
Department of Astronomy and Astrophysics, 
University of Toronto\\
P.O.Box 360, Richmond Hill, Ontario, Canada L4C 4Y6}
\email{(mochnacki,rucinski)@astro.utoronto.ca}

\begin{abstract}
We observed a field in the disk of the LMC on two 
consecutive nights in search of rapid variable stars. 
We have found two pulsating stars of type RRab and $\delta$ Sct, 
and four binary stars, among the latter one sdB or CV below
the LMC blue Main Sequence and three very close binary systems 
on the MS. At least one of the MS binaries, 
and possibly all three, are the first solar-type 
(W~UMa-type) contact binaries to be
detected in any extragalactic system and observed to obey the 
same $M_V=M_V(\log P, B-V)$ calibration as the Galactic 
systems. Given the selection effects due to small 
amplitudes at faint magnitudes, the frequency of 
such binaries in the disk of the LMC with its large
spread in population ages is not inconsistent 
with that in the disk of our Galaxy, and 
contrasts with the lack of binaries found in earlier
observations of the much younger LMC cluster LW55.
\end{abstract}


\section{INTRODUCTION}
\label{sec1}

Very little is known about short time scale ($<1$ day) 
variability of stars in the Magellanic Clouds.
Only recently, the availability of large telescopes located at 
excellent sites has made it possible to consider time-domain
monitoring of the stellar population at 
brightness levels reaching and beyond the levels of the
Turn-Off Point of the oldest stellar population in LMC
at $V \simeq 20.5$.

The current study addresses the detection and characterization 
of contact binary stars in a typical LMC field
1.7 degrees from the center of the LMC. This field was selected
taking guidance from an HST study of the stellar population in
the LMC \citep{SH2002},
where it was called Disk-1. Availability of the
archival HST images was one of the reasons for this study
as it permitted us to check for stellar blends and assure
better consistency of our results. The field is
characterized by a constant star formation rate from
the advanced age of 7.5 to 15 Gyr until
recently. In a companion study on short time-scale
variability in the LMC \citep{KR2003}, the studied field 
was dominated by a population of the open cluster LW55 
with an age of about 1.5 Gyr, in the presence of
an underlying older population with an 
age about 4 Gyr or more. We did not find any
short-period binaries in or around LW55, but only several
short-period pulsating stars instead.

We present the observations and show the color --
magnitude diagram for stars in the field in 
Section~\ref{obs}. The results of
the search for short-period variable stars are given in 
Section~\ref{var}. Section~\ref{concl} concludes 
the paper and summarizes the results.

\section{Observations}
\label{obs}

\subsection{Instruments and observing conditions}

We used the Magellan/Baade 6.5m telescope with the
TEK5 CCD $2K \times 2K$ camera which had a focal scale 
of 0.069 arcsec/pixel.
The field of view was $137 \times 137$ arcsec square.
The images were binned by 2 pixels in both directions
before extraction of photometry, with the resulting scale
of 0.138 arcsec/pixel. The median seeing during our 
run was 0.94 arcsec in the $V$ filter and 1.09 arcsec 
in the $B$ filter. The binned images significantly 
oversampled the Point Spread Function (PSF) even in the
cases of the best seeing.
The search for variability was done mainly with 
the $V$ filter (47 images) but also with the $B$ filter
(12 images). All $V$-filter exposures were 600~s while all 
$B$-filter ones were 900~s.

The observations were made on two nights, 2002 January 4/5 
and 5/6. We conducted 6.2 hours of variability monitoring 
on the first night and 7.3 hours on the second night.  
The nights can be characterized as gray time, with the 
fraction of the illuminated area of the Moon disc
of 61\% and 50\%, respectively.
 
The image processing was identical to that 
described in \citet{KR2003}.
The initial processing of the images was done with 
standard procedures from 
IRAF\footnote{IRAF is distributed by the 
National Optical Astronomy Observatories, 
which are operated by the Association of Universities for Research 
in Astronomy, Inc., under cooperative agreement with the NSF}
with a combination of the dome 
and sky flats used for the CCD response flat-field corrections.

The astrometric calibration was based on 48 reference stars 
from the USNO-B catalogue \citep{USNO-B}.
The random errors for the calibration do not
exceed 0.5 arcsec, based on the recovered coordinates
of the USNO-B stars.

\subsection{The field Disk-1}

\begin{figure}

\plotone{fig1}
\figcaption[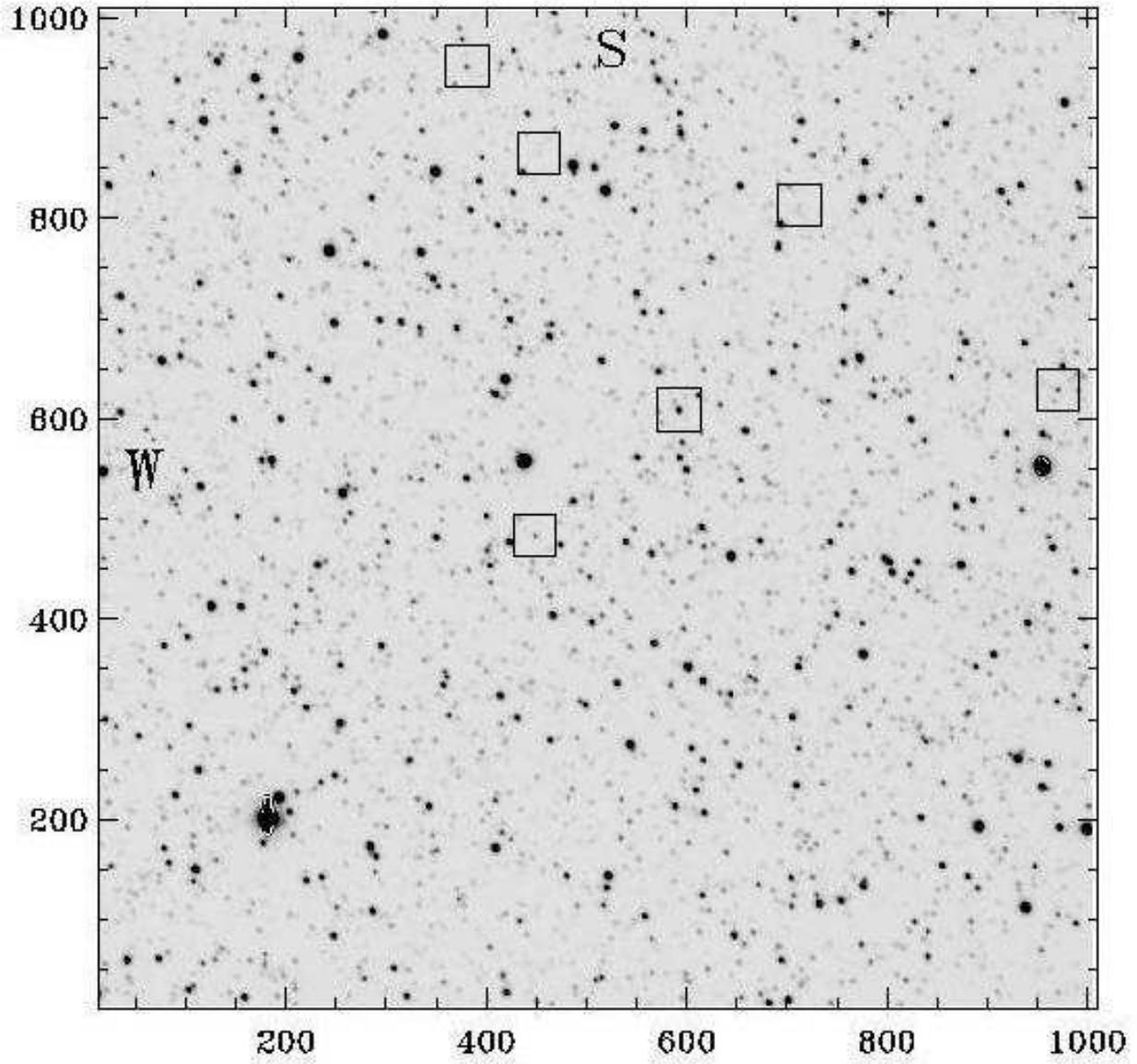] {\label{fig1}
The chart of the field in the pixel coordinates, the same as in
Table~\ref{tab1}. The variable stars can be found using the boxes
marked in the figure, the $X,Y$ coordinates given in
Table~\ref{tab2} and the individual charts
in Figure~\ref{fig8}.
}
\end{figure}

Figure~\ref{fig1} shows the sky image in the pixel space (after
the $2 \times 2$ binning), with pixels 0.138 arcsec
in size and with a total field of view of 
$2.3 \times 2.3$ arcmin square. The field in Figure~\ref{fig1} 
is oriented with East to left and North down. 
The $X$-coordinate runs W to E, while the $Y$-coordinate runs
N to S. 

\subsection{Photometric calibration}
\label{calib}

The photometric calibrations were done using three of Stetson's 
standard star fields \citep{stetson2}: 2 stars in NGC~1866,
6 stars in E4--108 and 15 stars in NGC~2682 (M67). 
The fields were observed with air masses ranging from 1.13 to 1.77. 
The extinction coefficients as well as the
color terms and zero points of the transformations from
the instrumental to the standard $BV$ system were
determined from observations of all 23 standards stars.
The adopted formulae were:
\begin{eqnarray*}
v & = & V - 1.050 + 0.021\times (B-V) + 0.089 \times (X-1.25) \\
b & = & B - 0.702 - 0.142\times (B-V) + 0.147 \times (X-1.25)
\end{eqnarray*}

\begin{figure}
\plotone{fig2}
\figcaption[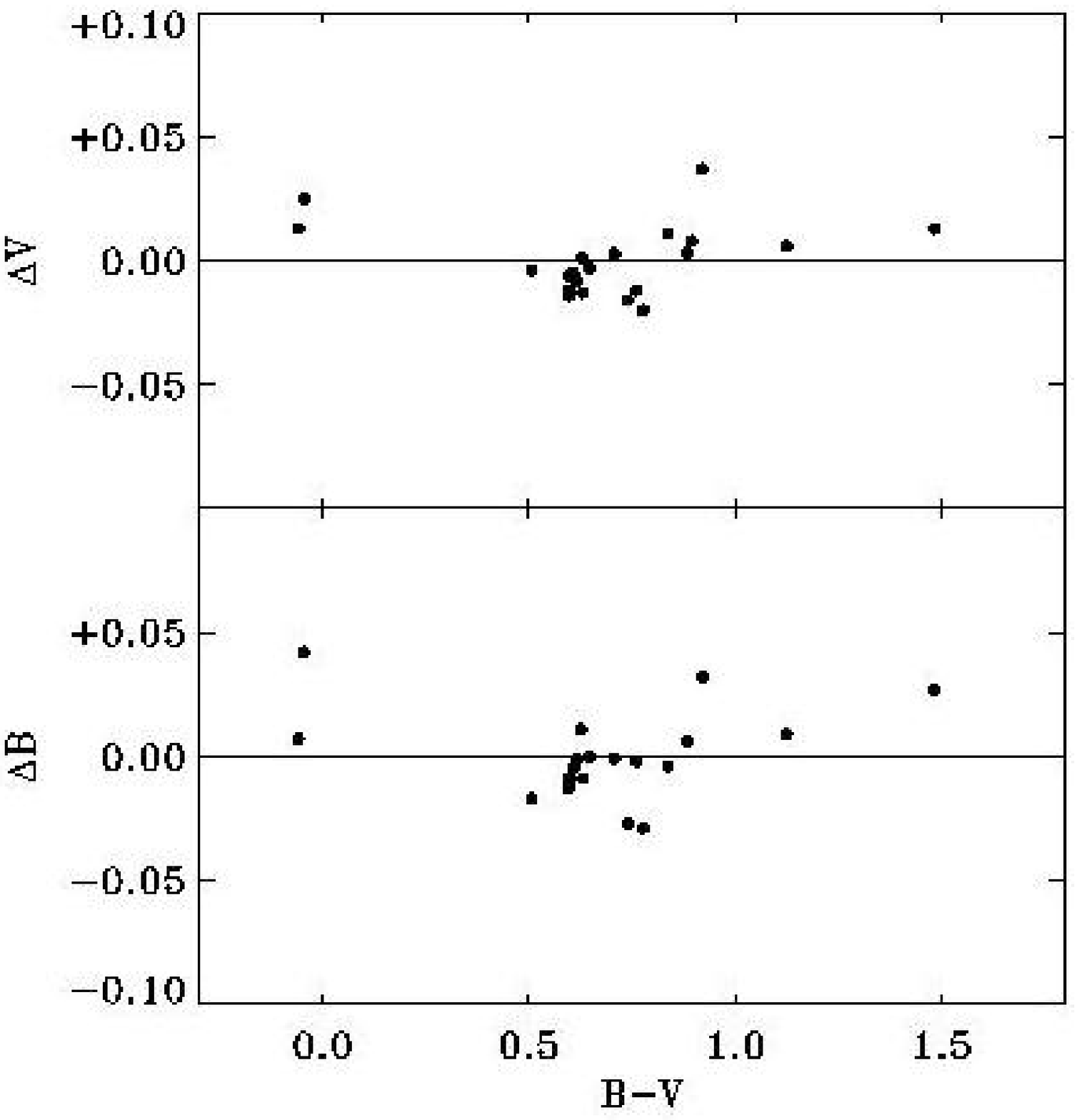] {\label{fig2}
Residuals for our photometry of the standard stars,
in the sense `our minus standard'' for $V$ and $B$ magnitudes.
}
\end{figure}

\noindent
$X$ is an air-mass and lower-case symbols 
denote instrumental magnitudes derived with the 
aperture photometry with the DAOPHOT program \citep{stetson}.
Figure~\ref{fig2} shows transformation
residuals for the standard stars.

\placetable{tab1}    

\begin{deluxetable}{rrrrrrrcc} 
\tabletypesize{\scriptsize}
\tablecolumns{9} 
\tablewidth{0pc} 
\tablenum{1}
\tablecaption{Photometric data for LMC Disk-1 \label{tab1}} 
\tablehead{ 
\colhead{\#} & \colhead{$X$} & \colhead{$Y$} & 
\colhead{RA[deg]} & \colhead{Dec[deg]} &
\colhead{$V$} & \colhead{$\sigma V$} & 
\colhead{$(B-V)$} & \colhead{$\sigma (B-V)$}
}
\startdata 
    1 &  16.43 & 841.77 & 77.92661 & $-$71.19987 & 22.409 &  0.019 &  0.580 &  0.031 \\
    2 &  16.95 & 736.51 & 77.92596 & $-$71.19575 & 22.363 &  0.008 &  0.339 &  0.016 \\
    3 &  17.06 & 175.30 & 77.92218 & $-$71.17359 & 22.357 &  0.006 &  0.350 &  0.012 \\
    4 &  17.85 & 548.81 & 77.92495 & $-$71.18848 & 17.787 &  0.004 &  0.831 &  0.008 \\
    5 &  17.86 & 423.94 & 77.92420 & $-$71.18362 & 22.716 &  0.008 &  0.498 &  0.017 \\
    6 &  18.43 &1002.93 & 77.92822 & $-$71.20638 & 21.848 &  0.035 &  0.399 &  0.056 \\
    7 &  18.73 & 760.67 & 77.92633 & $-$71.19669 & 22.535 &  0.007 &  0.429 &  0.013 \\
    8 &  18.87 & 461.99 & 77.92456 & $-$71.18511 & 21.001 &  0.004 &  0.172 &  0.008 \\
    9 &  19.23 &  88.02 & 77.92155 & $-$71.16988 & 23.817 &  0.015 &  0.533 &  0.037 \\
   10 &  19.99 & 662.28 & 77.92588 & $-$71.19287 & 23.804 &  0.021 &  0.494 &  0.041 \\
   11 &  20.07 &  56.83 & 77.92130 & $-$71.16853 & 21.416 &  0.007 &  0.389 &  0.016 \\
   12 &  20.12 & 488.31 & 77.92488 & $-$71.18614 & 22.957 &  0.008 &  0.435 &  0.016 \\
\enddata
\tablecomments{Table 1 is presented in its entirety in the electronic
edition of the Astronomical Journal. A portion is shown here for 
guidance regarding its form and content. 
The stars positions are given in the pixel $X$ and $Y$ 
coordinates of the reference image, as in Figure~\ref{fig1}, and in the 
equatorial coordinates in the J2000 system. The formal errors given here
can be used to compare relative uncertainties; the external estimates of
the errors are discussed in the text.
}
 
\end{deluxetable}

Table~\ref{tab1} contains the results of our
photometry for 4413 stars in the same pixel coordinate 
system of as shown in Figure~\ref{fig1}. The reference
frames in $V$ and $B$ were obtained by collation of,
respectively, 10 and 7 individual frames with the best seeing, 
in the way as described in detail in \citet{mochej02}.
The seeing in the reference images had
FWHM of 0.62 arcsec in $V$ and 0.74 arcsec in $B$.
The quoted errors in Table~\ref{tab1} do not reflect stellar
variability but are just the internal errors of the
PSF photometry on the reference images.

\subsection{Photometric uncertainties}

It is difficult to assess reliably errors of single-frame 
profile photometry in a crowded field such as Disk-1. 
Some formal errors, such as 
those returned by the DAOPHOT software, are often 
too optimistic as they tend to miss problems related 
to blending. 

We were not able to compare directly our
results with those of \citet{SH2002} 
which were discussed only in graphical and verbal form. 
However, we have analyzed the archival HST--WFPC2 data:
$V$: 2 $\times$ 500s, F555W, \#u4b10905r and \#u4b10906r; 
$I$: 2 $\times$ 300s, F814W, \#u4b10902r and \#u4b10904r; 
the pairs of images were used to remove cosmic rays.
We estimated deviations between our photometry
and the HST photometry, the latter obtained using the ``HSTphot'' 
software package \citep{Dol2002a,Dol2002b}.
While no obvious trends exist
in the $V$ magnitude differences for 963 common stars 
in the range $18 < V < 25$, there does exist an offset of
$\Delta V = -0.08 \pm 0.02$ (our results brighter).
This offset has not been applied to the photometric results
listed in Tables~\ref{tab1} and the variable stars
(Section~\ref{var}). We note that systematic uncertainties in
the WFPC2 photometry are estimated at about 0.05 mag.,
and may even reach 0.1 mag. \citep{Piotto2002}.

\begin{figure}
\plotone{fig3}
\figcaption[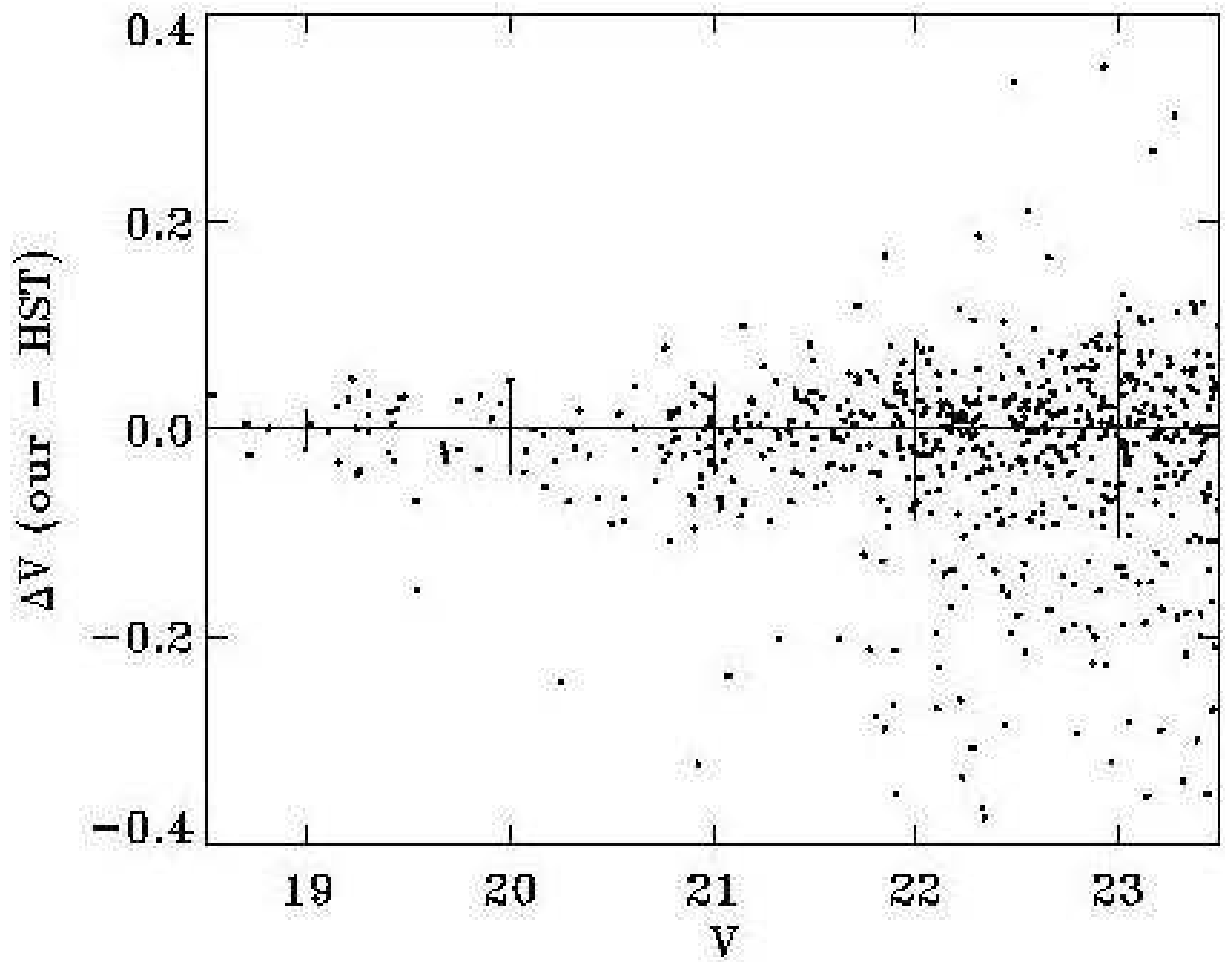] {\label{fig3}
$\Delta V$ differences between our photometry and HST photometry
(after allowance for the offset of $-0.08$ mag) can be
taken as a measure of random errors at different magnitude
levels. Estimates of the $rms$ errors at one magnitude
intervals are shown by vertical bars.
}
\end{figure}

To estimate random uncertainties, we compared 
our results with the HST results for individual stars.
The deviations for the sample of 963 common stars
are shown in Figure~\ref{fig3}. 
After accounting for the systematic shift of $-0.08$,
binning the deviations in one magnitude intervals, and with 
an assumption that errors from both sources add quadratically,
we estimate the $rms$ scatter at $\sigma V \simeq 0.02-0.04$ 
for for $V<21$; it increases to 0.09 at $V=22$ and to 
0.11 at $V=23$. 

Much more reliable than the above estimates are estimates of
errors in the search for variable stars, obtained by
comparison a large number of images analyzed using the image 
differencing technique. They are discussed in 
Section~\ref{var} in a discussion of amplitudes of
detectable variable stars.

\subsection{The color magnitude diagram}
\label{cmd}

\begin{figure}
\plotone{fig4}
\figcaption[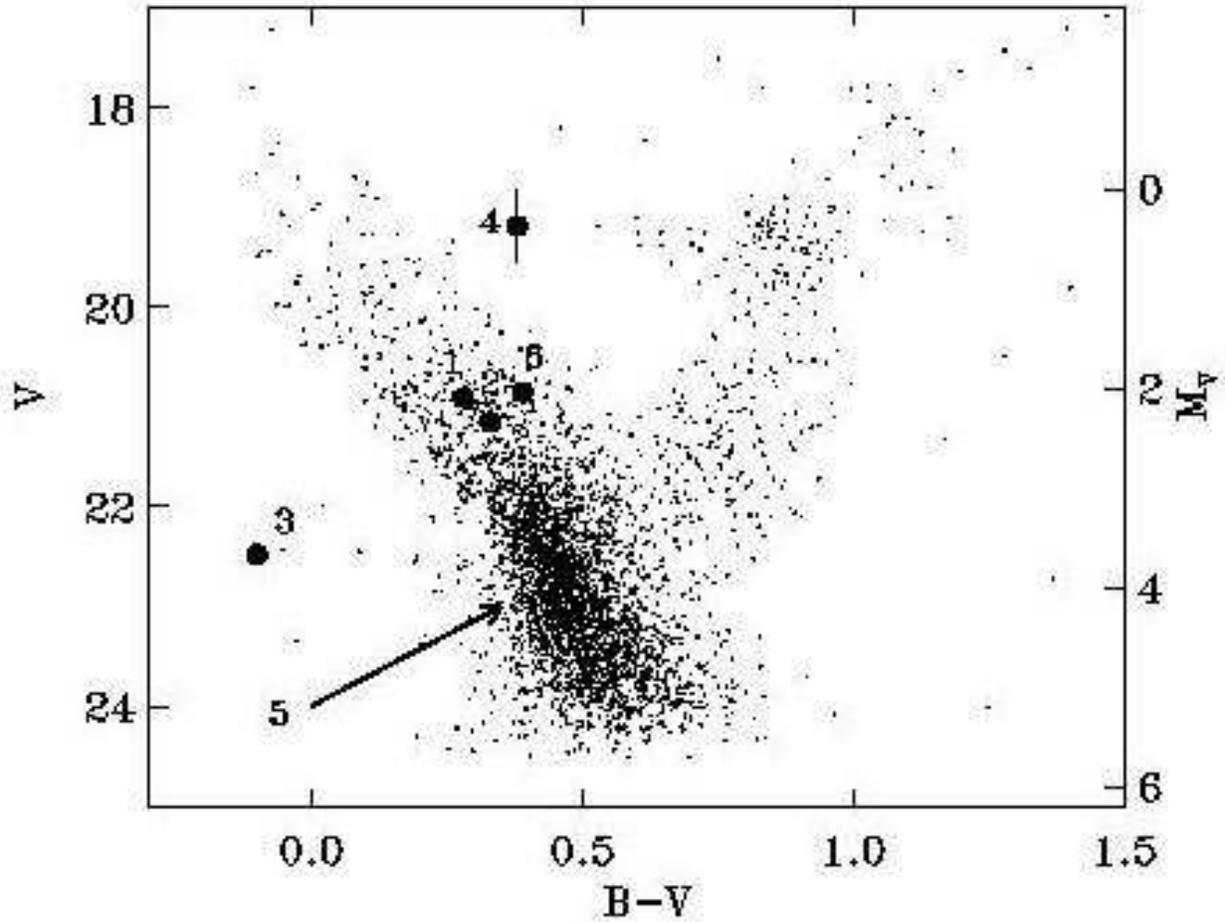] {\label{fig4}
The CMD for the Disk-1 field. The six variable stars are marked
with large filled circles at their
median magnitudes and color indices. The vertical bar for V4
shows its observed ranges of variability in $V$; for the
remaining variables the ranges are smaller than the symbol size.
The absolute magnitude scale on the right vertical
axis is based on the assumption of $(m-M)_0=18.5$ and
$E_{B-V}=0.10$.
}
\end{figure}

The photometric results for the whole field 
are plotted on the color--magnitude diagram
(CMD) in Figure~\ref{fig4}. As discussed by
\citet{SH2002}, the Disk-1 field contains populations of
different age with clear indications that the star
formation rate in this part of the LMC was 
constant over a long time. We can see the
red horizontal branch (at about $V \simeq 19.5$) 
and the Main Sequence turn-off point 
(at about $V \simeq 21$) of the old population 
as well as a sequence of young stars extending
to $V \simeq 18$.

\section{Variable stars in Disk-1}
\label{var}
\subsection{Techniques and errors}

A search for potential variable stars in the field Disk-1
was performed with the image subtraction package ISIS V2.1 
\citep{alard,alard2}.
Two methods were used to detect potential variable objects. 
First, we applied procedures which are included in the 
ISIS package and which are based on analysis of residual 
images. The second method relies on extraction -- 
still within the ISIS package -- of light curves 
for all stellar objects whose positions had been 
determined on template images 
with DAOPHOT/Allstar \citep{stetson}. 
Extracted light curves are subsequently
examined for the presence of any possible 
periodic variations with a suite
of programs based on the ``AoV'' algorithm \citep{alex}.

\begin{figure}
\plotone{fig5}
\figcaption[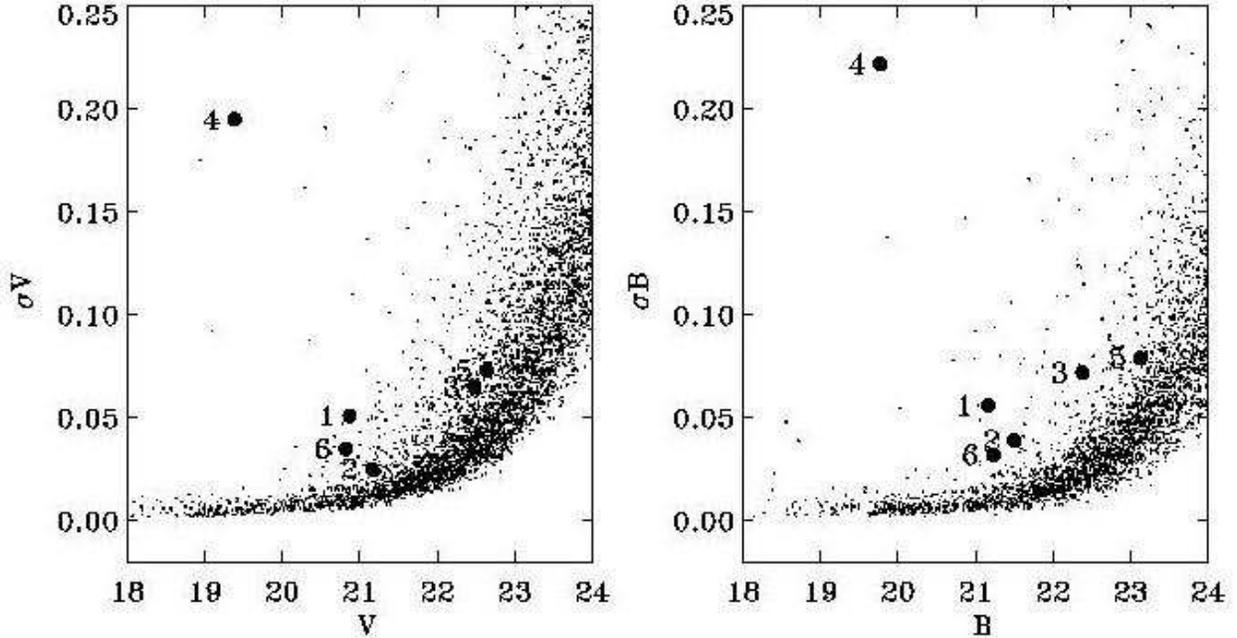] {\label{fig5}
The $rms$ errors of photometry based on the inter-comparison
of the individual images used for detection of variability.
The two panels show separately photometric errors in
$V$ and $B$ bands versus the respective magnitudes.
}
\end{figure}

Figure~\ref{fig5} shows the $rms$ versus the magnitude 
diagram for 6243 stars whose light curves were 
analyzed (this number is by about one half 
larger than the number of stars that went into the CMD). The 
$rms$ errors were derived after rejection of the 
two most extreme maximal and minimal data points from data for
each star. Not all points with large $\sigma$ in the figure
correspond to physical variables as automatic processing led to
inclusion of a few easily identifiable
stars with poor photometry, mostly
in the extended wings of over-exposed bright stars.
The figure suggests that assuming a $5 \times 
\sigma$ detection threshold, 
we should be able to detect variable stars with peak-to-peak
amplitudes of $\Delta V \simeq 0.05$ at $V = 21.5$ and 
$\Delta V \simeq 0.10$ at $V = 22.5$. 
It is encouraging to note that the data were taken 
during gray time. Clearly, even during 
gray time, one may use the Baade telescope to look for 
variable stars in the LMC beyond the turnoff region at  
$V \simeq 22$.

\subsection{Results of the variable star search}

We have detected six variable stars in the Disk-1 field. 
They are marked on the color-magnitude diagram of the
field in Figure~\ref{fig4} and are listed in Table~\ref{tab2}, 
where their CMD numbers, $X,Y$ and
equatorial J2000 coordinates, the maximum
and minimum $V$ and median values of $(B-V)$ are given. 
The listed  values of $(B-V)$ are  
uncertain because of the non-simultaneous nature of 
our $B$ and $V$ observations and because 
the exposure times were rather long
(15 minutes in $B$). The $B-V$ data given in Table~\ref{tab2}
are slightly different than the ones in Table~\ref{tab1} due to 
stellar variability along with differences in how 
the mean photometric values were determined.

\placetable{tab2}    
\noindent
\begin{deluxetable}{rrrrrrcccccc} 
\tabletypesize{\scriptsize}
\tablecolumns{12} 
\tablewidth{0pc} 
\tablenum{2}
\tablecaption{Variable stars in Disk-1 \label{tab2}} 
\tablehead{ 
\colhead{V} & \colhead{\#} & \colhead{$X$} & \colhead{$Y$} 
& \colhead{$V_{max}$} & \colhead{$V_{min}$} & \colhead{$(B-V)$} 
& \colhead{RA(hh:mm:ss)} & \colhead{Dec(dd:mm:ss)}
& \colhead{Type}
& \colhead{Period(d)} & \colhead{$T_0$}
}
\startdata 
1 & 1662 & 381.82 & 953.11 & 20.84 & 20.97 & \phs 0.28 & 5:11:53.10 & $-$71:12:14.6 
& EA/EW:  & 0.491(1): & 79.715 \\
2 & 1984 & 449.61 & 484.19 & 21.12 & 21.18 & \phs 0.33 & 5:11:54.51 & $-$71:11:09.4
& $\delta$~Sct & 0.0675(1) & 79.568 \\
3 & 1999 & 453.81 & 866.78 & 22.39 & 22.57 & $-$0.10   & 5:11:55.02 & $-$71:12:02.0 
& sdB/CV & 0.2607(2) & 80.67  \\
4 & 2618 & 593.36 & 609.87 & $<$18.80 & 19.56 & \phs 0.38 & 5:11:58.71 & $-$71:11:25.8 
& RRab & 0.538: & \\ 
5 & 3175 & 713.90 & 814.92 & 22.57 & 22.78 & \phs 0.47 & 5:12:02.35 & $-$71:11:53.2 
& EW & 0.3108(1) & 80.625 \\
6 & 4284 & 971.90 & 630.27 & 20.79 & 20.91 & \phs 0.39 & 5:12:09.52 & $-$71:11:26.2 
& EA/EW: & 0.469(1): & 80.69 \\
\enddata
\tablecomments{The variable star numbers, as used in the text,
are given in the first column, while identifications in Table~\ref{tab1}
are in the second column.
The full names which conform to the International
Astronomical Union recommendations are ``Disk1--LCO--V...''. 
The J2000 equatorial coordinates were obtained through
an astrometric frame solution using positions of 48 stars 
from the USNO-B catalogue; this solution reproduces the
J2000 coordinates of these stars with residuals not exceeding 
0.5 arcsec in RA and Dec. 
Note that the $V$ and $B-V$ data differ slightly between this table
and Table~\ref{tab1} due to different photometric methods used; the
differences can be taken as indication of external
uncertainties combined with the genuine variability.
}
\end{deluxetable}

\begin{figure}
\plotone{fig6}
\figcaption[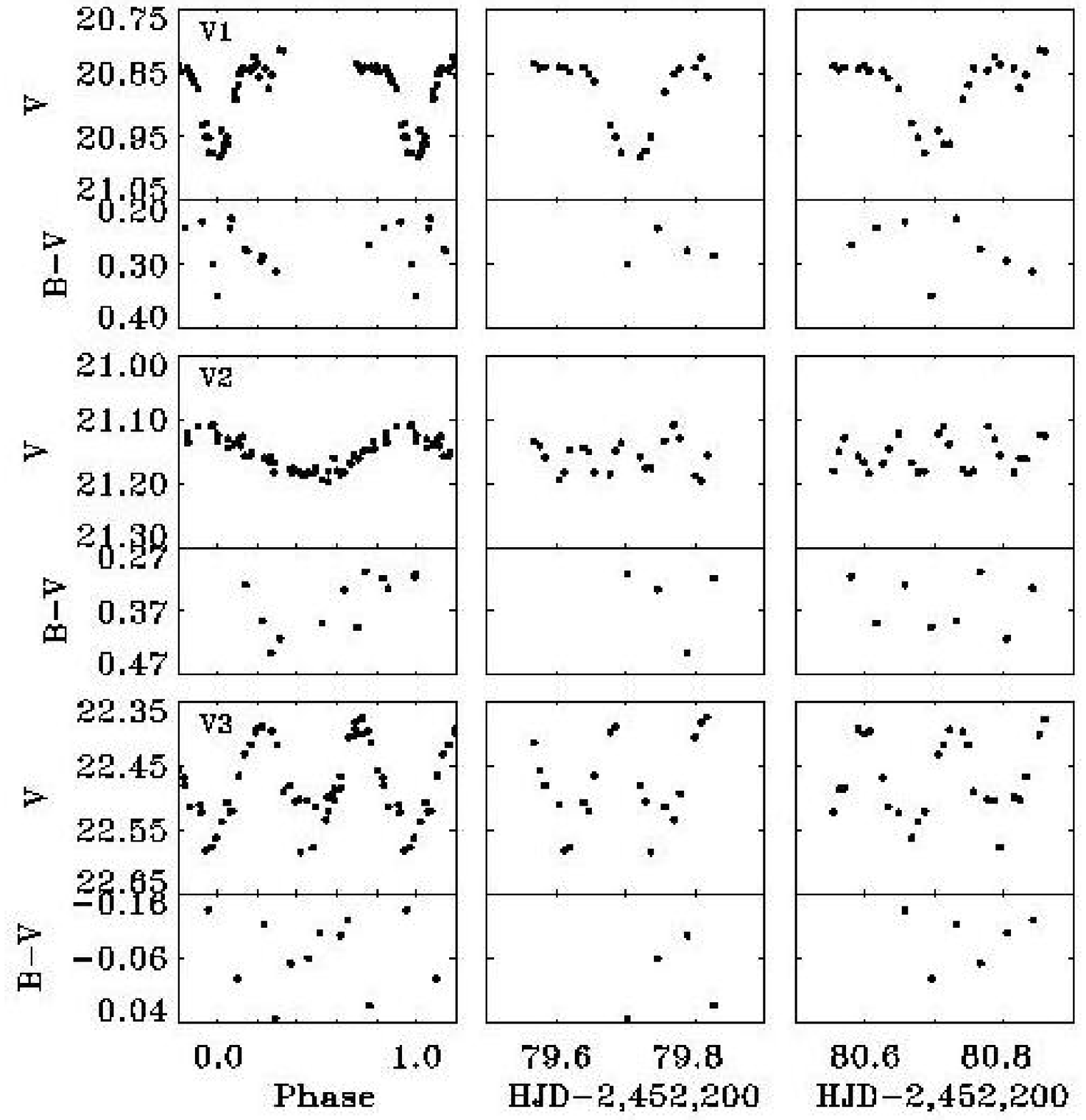] {\label{fig6}
The light and color index curves for variables V1 -- V3.
For each star,
the left panel shows the phased data with $T_0$ and the
period as given in Table~\ref{tab2}, while the two right
panels show the individual nightly data. The vertical scale
has the same range of $\Delta V = 0.3$ and
$\Delta (B-V)$ for all stars, but
the magnitude and color index levels are different.
}
\end{figure}

\begin{figure}
\plotone{fig7}
\figcaption[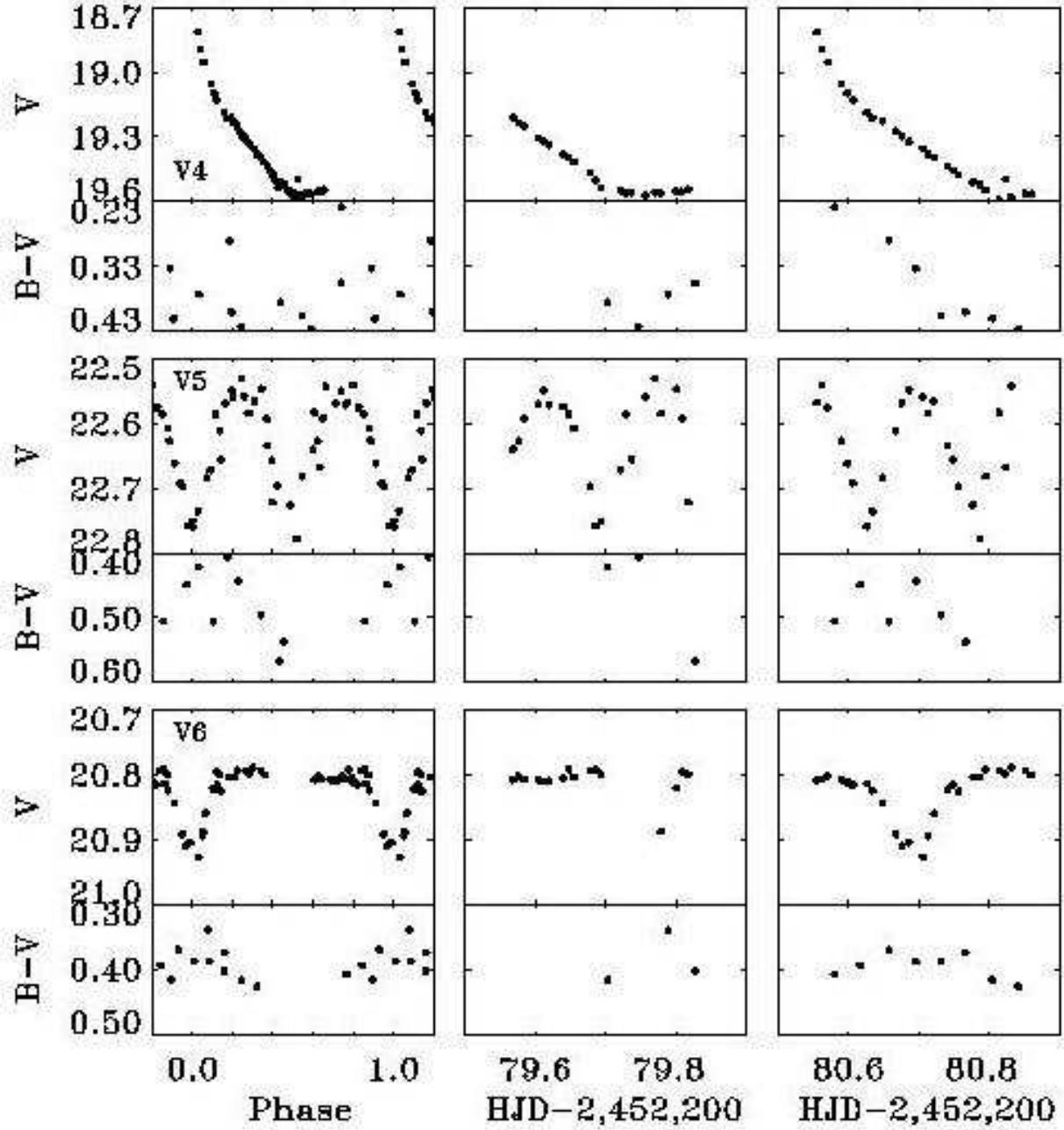] {\label{fig7}
The same as in Figure~\ref{fig6}, but for the variables
V4 -- V6. Note that for V4, the magnitude scale is expanded
3 times relative to that for the remaining stars.
}

\end{figure}

\begin{figure}
\plotone{fig8}
\figcaption[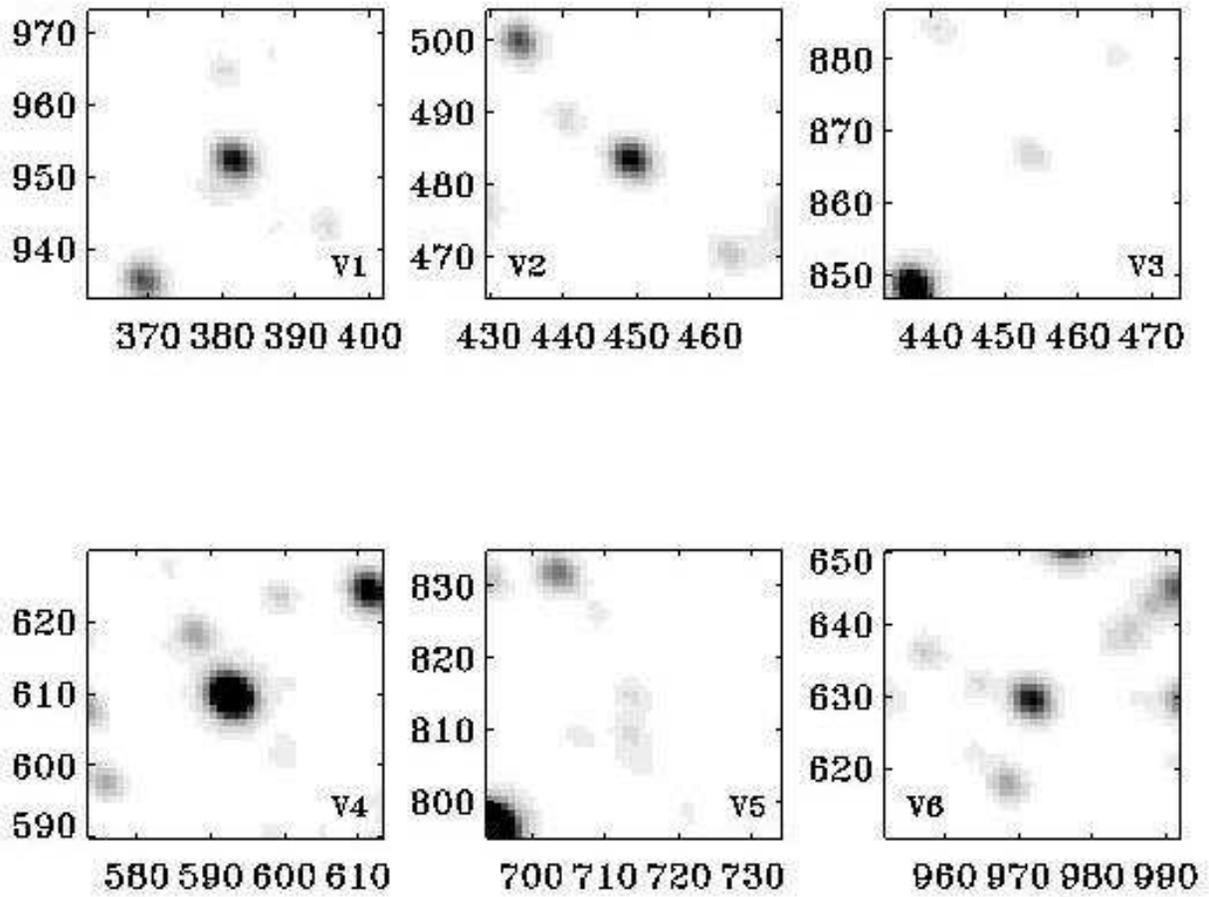] {\label{fig8}
The finding charts of the variable stars showing
$40 \times 40$ pixels of our reference image (Figure~\ref{fig1})
centered on each of the variable stars.
}
\end{figure}

The light curves of the variables\footnote{The tables
of individual $V$ and $B$ magnitudes and interpolated
$B-V$ are available from:
http://astro.utoronto.ca/rucinski/LMC-Disk1/Var*\_Disk1.dat}
are shown in Figures~\ref{fig6} and \ref{fig7}, 
while the finding charts in pixel coordinates of 
Figure~\ref{fig1} are shown in Figure~\ref{fig8}. 
Figures~\ref{fig6} and \ref{fig7} give the $V$ and
$B-V$ curves, the latter for the interpolated 
moments of the $B$ observations. We do not discuss the
$B$ magnitude light curves in this paper because of their
low precision.

We inspected the archival WFPC2 images (particularly
the image WFPC2ASN U4B10905B) and found that all our
detected variables are free from blending. We also
found that our RA/Dec coordinates of the variables
agree with the WCS of the WFPC2 image to 0.6 arcsec.
 
Obviously, due to the short duration of the program,
we could detect only short time scale variables. On the
other hand, we could probe relatively deeply and could
detect variables among stars of very moderate brightness 
reaching as far down the main sequence as solar stars, 
in the LMC at $V \simeq 22 - 23$. The limited scope of the
project is fully confirmed by the characteristics of
the detected variables: As we can see in Fig.~\ref{fig4},
four of the stars are located in the Main Sequence of
the LMC; these are V1, V2, V5, V6.  One of them, V2, is
a $\delta$~Scuti variable, while the remaining ones are short
period binary systems. V1, V2 and V6 are located in the
turn-off region of the old population Main Sequence within 
$+0.28 < B-V < +0.39$ and $V \simeq 21$.
We also detected one RR~Lyr-type (RRab)
variable and one very blue variable below the
Main Sequence.
We comment on the individual objects below. 
\begin{itemize}
\item V1 (\#1662) -- A short period eclipsing binary
with an amplitude of light variations of $\Delta V \simeq 0.15$,
period $P = 0.49$ d and $B-V=0.28$. 
Although the light curve may suggest a close, but detached
binary, $M_V = 2.0$ predicted from $(m-M)_0=18.5$ and
$E_{B-V}=0.1$ perfectly agrees with  the contact-binary
calibration of \citet{RD97}(see also \citet{ruc00}), 
$M_V(cal)=2.0$.
\item V2 (\#1984) -- A short-period (0.0675 d) $\delta$~Sct or
SX~Phe pulsating star with $\Delta V = 0.06$ at $V = 21.15$.
The light curve is relatively well
defined thanks to the folding of several periods.
\item V3(\#1999) -- An interesting blue star well below
the Main Sequence at $V \simeq 22.45$ and $B-V \simeq -0.1$
(independent photometry on the reference image of 
Section~\ref{cmd} gives $B-V \simeq -0.05$). Variability
is rapid; if it is a binary, as individual nightly
observations suggest, then the period is 0.2607 day, but
the period may be actually 1/2 of this value.
We verified that the blue color of the star is 
not due to blending, which is possible in such 
a crowded field as Disk-1
because the archival WFPC2/HST images mentioned 
above clearly show that V3 is indeed very blue.
The observed luminosity of V3 is consistent with 
a relatively bright Cataclysmic Variable in the LMC. 
The variable is isolated and bright enough that 
low-resolution spectra could be obtained from the ground. 
\item V4 (\#2618) -- This is an RRab pulsating star with 
well defined light variations, but an incomplete light curve. 
We assumed that two cycles elapsed between the two nights,
P=0.538 d. 
Since we have not captured the light maximum, we have been
unable to determine the initial (maximum light)
epoch $T_0$ for this star;
the maximum brightness is possibly above $V = 18.8$.
Because the light curve is incomplete, we cannot relate
photometric properties of the star to the horizontal
branch of the LMC, but the properties are certainly 
consistent.
\item V5 (\#3175) -- A rather typical contact binary 
well within the Main Sequence of the LMC, some +2
magnitudes below the Turn-Off Point, at $V_{max}=22.57$. 
Assuming $B-V = 0.47$, $E_{B-V} =0.1$ and $P=0.3108$ d, the
calibration of \citet{RD97} predicts $M_V (cal) = 3.5$;
directly, with $(m-M)_0 = 18.5$, one obtains
$M_V=3.8$. This is consistent within the 
current uncertainties. V5 is the first certain
W~UMa-type binary identified in other galaxy.
\item V6 (\#4284) -- A close eclipsing binary 
with $V_{max} \simeq 20.8$, $B-V = 0.39$
and the period of 0.469 d (assuming the
observations cover two cycles).
From the distance modulus, $M_V = 2.0$, while --
assuming that the binary is really a contact one -- 
the calibration of \citet{RD97} gives $M_V (cal) = 2.4$;
the light curve does not look like that of a contact 
binary, however.
\end{itemize}

\section{Summary and conclusions}
\label{concl}

We have photometrically observed a region of the 
Large Magellanic Cloud -- called ``Disk-1'' by \citet{SH2002} -- 
on two consecutive nights in search of short time scale variable
stars. The time monitoring led to the discovery of four
short-period eclipsing binaries and of two pulsating stars
(RRab and $\delta$~Sct). The results clearly show that
solar-type stars are accessible for such monitoring in
the LMC down to $V \simeq 22.5 - 23$. 

It appears to be significant that the field Disk-1 
contains short period binaries. We discovered four
such binaries, in contrast to the results for the 
LMC field of LW55 \citep{KR2003} 
which was observed in a practically
identical way and where we detected only very low
amplitude, pulsating variables of $\delta$~Scuti, SX~Phoenicis
or $\gamma$~Doradus type, and no binaries. 
While small-number statistical
fluctuation is still a possibility, the reason may be
in the uniform distribution of stellar ages in the
Disk-1 field \citep{SH2002}, contrasted with the
population dominated by  LW55 itself with an age of
1.5 Gyr.

Obviously, the number of short-period binaries is still small,
4 among 6243 monitored including possibly 3 contact binaries,
compared with the Galactic field where typically one among
500 FGK dwarfs is expected to be a contact binary
\citep{ruc02}.
However, all binaries detected by us have moderately
large amplitudes, $\Delta V > 0.1 - 0.15$, 
which may result from selection effects operating at the faint
levels of $V \simeq 24$ in a very crowded field; 
we simply could not detect low amplitude binaries.  
Numbers of close binaries apparently increase 
rapidly with decreasing amplitude of light
variations \citep{ruc01}. This is confirmed  by the 
complete ($\Delta V > 0.05$) sample of bright, 
short-period binaries in the solar neighbourhood, mostly from the Hipparcos catalog, 
with $V < 7.5$ \citep{ruc02}.
In this sample, only about 1/3 of all binaries 
have amplitudes $\Delta V > 0.10$ 
and almost 1/2 of them have amplitudes 
$\Delta V < 0.15$. Taking this 
amplitude selection effect into account,
the discrepancy between the observed 3 and the
expected 8 contact binaries does not appear to
be significant, assuming that about 2/3 of the 
monitored stars were FGK dwarfs.

\acknowledgements

\noindent
We thank the reviewer Dr.\ Wayne Osborn for very careful checking
of our paper and for several useful suggestions.

Research support from the Ministry of Scientific 
Research and Informational Technology, Poland to JK 
(grant 1~P03D~001~28) and
from the Natural Sciences and Engineering Council of Canada
to SWM and SMR is acknowledged here with gratitude. 

This work is based on observations with the NASA/ESA Hubble 
Space Telescope, obtained from the Data Archive at the 
Space Telescope Science Institute, which is operated by 
the Association of Universities for Research in Astronomy, 
Inc., under NASA contract NAS 5-26555. These observations 
are associated with program \#7382

\clearpage                    

\noindent

\end{document}